\documentclass[conference]{IEEEtran}
\usepackage{graphicx}
   \graphicspath{{./Figures/}}

\usepackage{amsmath}    
\usepackage{algorithm}
\usepackage{algorithmic}
\usepackage{multirow}
\usepackage{makecell}
\usepackage{cite}
\usepackage[numbers,sort&compress]{natbib}
\usepackage{svg}
\usepackage{color}
\DeclareGraphicsExtensions{.pdf,.jpeg,.png,.fig, .emf}
   \usepackage{subfigure}
    \usepackage{epstopdf}
    \usepackage{epsfig}
    \usepackage{subfigure}
    \usepackage{epstopdf}
    \usepackage{epsfig}
\usepackage{booktabs}

\begin{document}
%
\title{Low Power and Temperature-Resilient Compute-In-Memory Based on Subthreshold-FeFET}

 \author{\small{}
         Yifei~Zhou$^1$, Xuchu~Huang$^1$, Jianyi~Yang$^1$, Kai~Ni$^2$, Hussam~Amrouch$^3$, Cheng~Zhuo$^{1,4,*}$, and Xunzhao~Yin$^{1,4,*}$\vspace{0.5em}\\
    $^1$Zhejiang University, Hangzhou, China; $^2$Department of Electrical Engineering, University of Notre Dame, Notre Dame, USA\\
    $^3$Chair of AI Processor Design, Technical University of Munich; TUM School of Computation, Information \\ and Technology; Munich Institute of Robotics and Machine Intelligence, Munich, Germany\\
    $^4$Key Laboratory of Collaborative Sensing and Autonomous Unmanned Systems of Zhejiang Province, Hangzhou, China \\
    $^*$Corresponding authors, email: \{czhuo, xzyin1\}@zju.edu.cn
}

\maketitle
\pagestyle{empty}

\begin{abstract}
Compute-in-memory (CiM) is a promising solution for addressing the challenges of artificial intelligence (AI) and the Internet of Things (IoT) hardware such as "memory wall" issue.
Specifically, CiM employing nonvolatile memory (NVM) devices in a crossbar structure can efficiently accelerate multiply-accumulation (MAC) computation, a crucial operator in neural networks among various AI models.
Low power CiM designs are thus highly desired for further energy efficiency optimization on AI models.
Ferroelectric FET (FeFET), an emerging  device, is attractive for building ultra-low power CiM array due to CMOS compatibility, high $I_{ON}$/$I_{OFF}$ ratio, etc.
Recent studies have explored FeFET based CiM designs that achieve low  power consumption. 
Nevertheless, subthreshold-operated FeFETs, where the operating voltages are scaled down to the subthreshold region to reduce array power consumption, are particularly vulnerable to temperature drift,
leading to accuracy degradation.
To address this challenge, we propose a temperature-resilient 2T-1FeFET CiM  design that performs MAC operations reliably at subthreahold region from 0℃ to 85℃, while consuming ultra-low power. 
Benchmarked against the VGG neural network architecture running the CIFAR-10 dataset, the proposed 2T-1FeFET CiM design achieves $89.45\%$ CIFAR-10 test accuracy. Compared to previous FeFET based CiM designs, it exhibits immunity to temperature drift at an 8-bit wordlength scale, and achieves better energy efficiency with 2866 TOPS/W.
\end{abstract}

%



\section{Introduction}
\label{sec:introduction}
With the rapid development of artificial intelligence (AI) and  Internet of things (IoT), there is an increasing demand for 
processing massive amount of data, thus calling for computing devices with low power consumption and read/write latencies \cite{deng2019energy}. Unfortunately, data-intensive applications are challenging to address with von Neumann architectures because of the  “memory wall” problem \cite{Wulf_1995}, where frequent data transfers lead to high power consumption. 
To overcome such challenges, computing-in-memory (CiM) has been proposed as a potential solution for energy-efficient AI and IoT hardware designs \cite{yin2023ultracompact, wei2023imga, eldebiky2023correctnet,  shou2023see, yan2022computing}, as 
it allows highly parallel arithmetic or logical calculations within the memory, which greatly increases the computing efficiency by eliminating the data transfers \cite{chen2022accelerating, yan2022swim, yin2022ferroelectric, wang2021triangle, huang2023fefet, yan2023improving}.

A number of non-volatile memory (NVM) devices have been considered for CiM designs,
such as resistive random access memeory (ReRAM) \cite{Yu_2021}, phase-change memory (PCM) \cite{Khaddam_2021}, ferroelectric FET (FeFET) \cite{ni2019ferroelectric, Soliman_2020,cai2022energy, Sk_2023, liu2022cosime, liu2023reconfigurable, xu2023challenges}, etc. 
Among these NVMs, FeFET offers advantages such as CMOS compatibility, ultra-low leakage current,  high $I_{ON}$/$I_{OFF}$ ratio and low write/read power. Prior works have reported FeFET based CiM designs for multiply-accumulation (MAC)  operation, the core operator in AI models.
Various FeFET based cells have been proposed to optimize the power consumption, speed, area and robustness \cite{Soliman_2020,Sk_2023,Lu_2022}, most of which operate in the saturation region of FeFETs.
That said, continuous optimizations on the energy efficiency of CiM solutions  are still desirable, 
given the  billion level number of MAC operations required for neural network processing. 
This is where subthreshold computing comes into play. 
By scaling the operating voltage of FeFET to the subthreshold region (subthreshold-FeFET), FeFET based CiM array can achieve further reductions in power consumption compared to those operating in the saturation region.

On the other hand, the increased computation density in a compact area leads to higher power density and temperature elevation \cite{Meng_2022}. Also, the operating temperature will be affected by environmental conditions. 
The varying temperatures that may change the circuit operation states are normally neglected by FeFET based CiM designers, assuming a thermostatic condition.
Unfortunately, FeFET is a highly temperature-sensitive device, and even more sensitive in the subthreshold region \cite{amrouch_fefet_device_temperature}. 
As temperature changes, the output of FeFET will change significantly, resulting in the error of circuits. Therefore, merely scaling the operating voltage to the subthreshold region is not sufficient, necessitating temperature-resilient FeFET based design. 


In this paper, 
to further scale down the power consumption of FeFET based CiM design, and to solve the temperature drift problem associated with the FeFET operating conditions, 
for the first time, we propose a 2T-1FeFET cell design that almost eliminates the impact of temperature ranging from 0℃ to 85℃ on the subthreshold-FeFET based CiM array, while achieving ultra-low power consumption. 
The operation and the temperature resilience of the proposed design have been illustrated and validated at both cell and array levels.
Evaluation results of the proposed design on VGG nerual network for Cifar-10 datasets demonstrate evident power consumption improvements with great resilience to temperature drift compared to other FeFET based CiM designs. 




The paper is organized as follows: Sec. \ref{sec:background} provides a  review of FeFET and subthreshold computation, and points out the design challenges. Sec. \ref{sec:proposedwork} analyzes impacts of temperature drift and proposes a temperature-resilient 2T-1FeFET cell. Sec. \ref{sec:eval}   evaluates the performance of our design.  Sec. \ref{sec:conclusion} concludes.

\section{Background}
\label{sec:background}

In this section, we review the FeFET device and the model capturing the temperature dependency of FeFET, introduce the subthreshold CiM design, and analyze the design challenges of temperature-resilient subthreshold-FeFET based CiM.

\subsection{FeFET Basics}
\label{sec:existing_work}

FeFETs, which utilize $HfO_2$ as the ferroelectric dielectric, have emerged as a competitive option for embedded NVM due to their ultra-low leakage current, high $I_{ON}$/$I_{OFF}$ ratio, voltage driven mechanisms and CMOS compatibility \cite{hu2021memory}. 
FeFETs allow for the switching of ferroelectric polarization within the gate layer by applying positive or negative gate pulses, which sets FeFET to the low-$V_{TH}$/high-$V_{TH}$ state \cite{khan2020future}, as shown in Fig. \ref{fig:templine}.
The stored data (i.e., '1' and '0') can then be read through the drain current (i.e., $I_{ON}$ and $I_{OFF}$) by applying a gate voltage between the low-$V_{TH}$ and high-$V_{TH}$.
Compared with other NVM devices, i.e., ReRAM \cite{Yu_2021}, PCM \cite{Khaddam_2021}, which require high power during the write due to large current, FeFETs provide superior energy efficiency due to the electric field driven write scheme
\cite{salahuddin2018era}.


Various models have been proposed to simulate FeFET devices in circuit designs, such as the negative capacitance FET (NCFET) based model \cite{aziz2016physics}, the multi-domain Preisach model \cite{ni2018A}, and the comprehensive Monte Carlo model \cite{deng2020acomprehensive}. However, these models lack consideration for temperature effects on devices, hindering the validation of functions and performance evaluation of FeFET based circuits in practical edge devices.
While some studies have explored the impact of temperature on FeFET devices \cite{amrouch_fefet_device_temperature, thomann2021reliability} and vulnerability of CiM architectures to device variations \cite{yan2022computing}, this work goes further by investigating and addressing the temperature effects on subthreshold-FeFET based CiM structures. To analyze the collective temperature effects on CiM at the circuit level, we utilize the experimentally calibrated Preisach FeFET compact model \cite{ni2018A} in conjunction with the Intel FinFET model.


\subsection{Subthreshold Computing Design}
\label{sec:design}

Scaling the operating voltage to subthreshold region rather than saturation region is a promising way to  optimize the energy efficiency, as the energy consumption is typically proportional to the conducting current and the voltage.
That said, subthreshold computation has only been studied in mature devices such as CMOS \cite{Loyez_2021}. These designs consume large area overhead and high power consumption, sacrificing the advantages of voltage scaling, and yet CiM designs based on subthreshold devices have not been studied so far.

\begin{figure}
    \centering
    \includegraphics[width=0.9\linewidth]{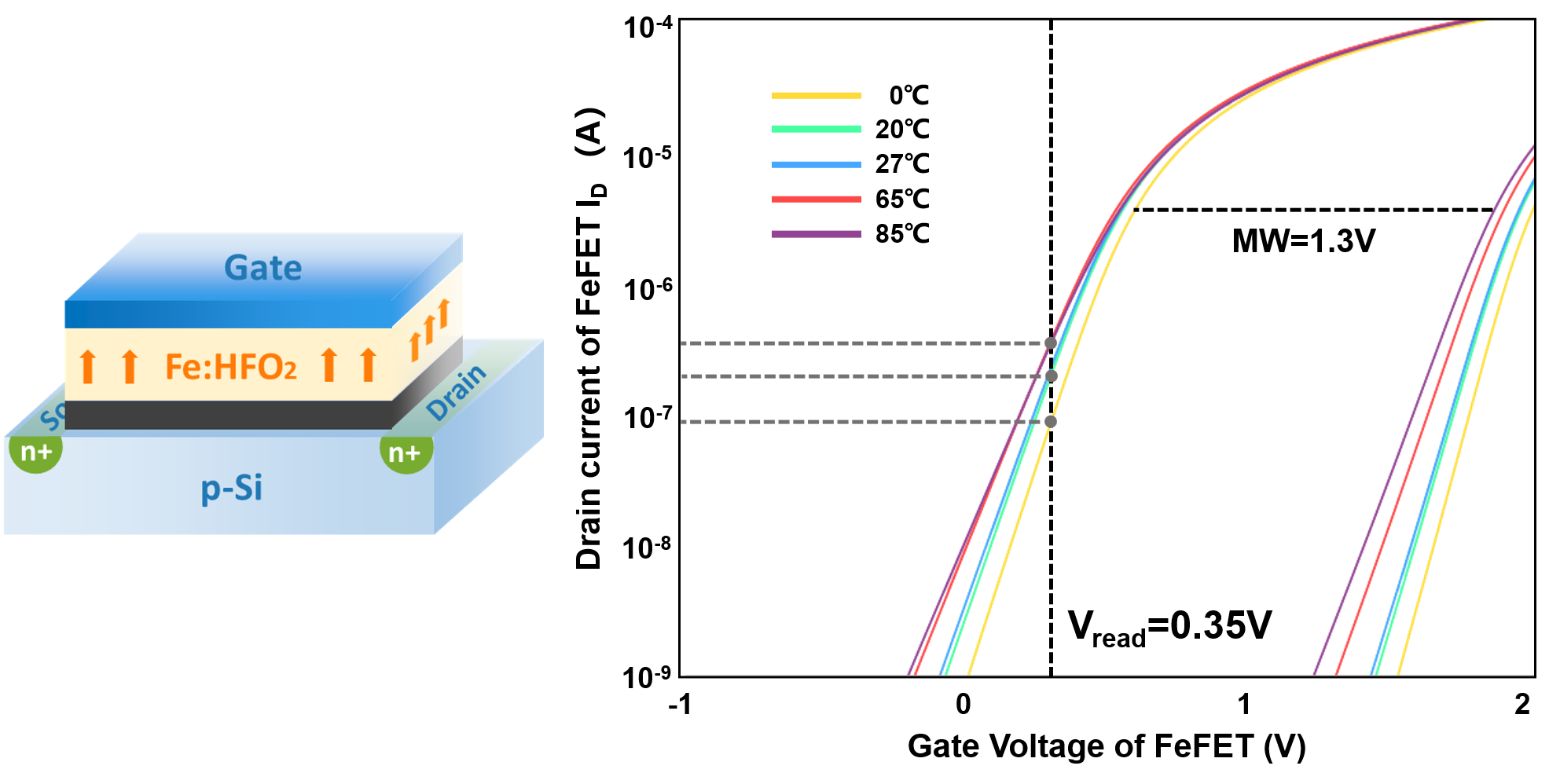}
    \vspace{-0.5em}
    \caption{FeFET structure and characteristic line at different temperatures and at two states (low-$V_{TH}$/high-$V_{TH}$). FeFETs working under the chosen $V_{read}=0.35V$ in our design are fully lying in the subthreshold region.}
    \label{fig:templine}
    \vspace{-1.5em}
\end{figure}

To address  the challenges faced by CMOS based subthreshold circuit design, in this work, for the first time, we propose a  FeFET based subthreshold CiM design for MAC operations. 
By leveraging the characteristics of FeFETs similar to MOSFET devices as shown in Fig. \ref{fig:templine}, subthreshold-FeFETs are acquired by applying scaled read voltage $V_{read}$ between the programmed memory window as shown in Fig. \ref{fig:templine}. 
As a result,  the current $I_{D}$ of FeFETs  is lowered down, achieving significant power  reduction.
However, similar to the CMOS based subthreshold computing, the subthreshold-FeFET  is highly  influenced by the temperature, as illustrated in Fig. \ref{fig:templine}. 
It can be seen that temperature changes have a stronger impact on the high-$V_{TH}$ state compared to the low-$V_{TH}$ state.
The detailed impact of the temperature on the subthreshold-FeFET based CiM design is analyzed in Sec. \ref{sec:proposedwork}.

\subsection{Subthreshold Operation vs. Temperature  Resilience}
\label{sec:contradiction}

Subthreshold-FeFET offers the advantage of reduced power consumption, while scaling voltages also makes the devices more susceptible to temperature drift, which deteriorates their performance or even functionality due to the exponential I-V relationship of FeFETs in the subthreshold region. 
Previous studies on CiM designs for MAC have already highlighted the challenges of achieving temperature-resilient design and subthreshold MAC operations. 
Combining both of these aspects further exacerbates the difficulty. Therefore, the challenge lies in finding an appropriate balance point between lower operating voltage and better resilience to temperature drift.



\section{Proposed Temperature-resilient 2T-1FeFET CiM Design}
\label{sec:proposedwork}

In this section, we study the impacts of temperature drift on existing FeFET CiM designs operating in the subthreshold region, and 
propose a novel cell structure that aims at improving temperature resilience and energy efficiency based on subthreshold-FeFET.

\subsection{Temperature Drift on Existing Designs}

\begin{figure}
     \centering
     \vspace{-0.5em}
     \includegraphics[width=0.8\columnwidth]{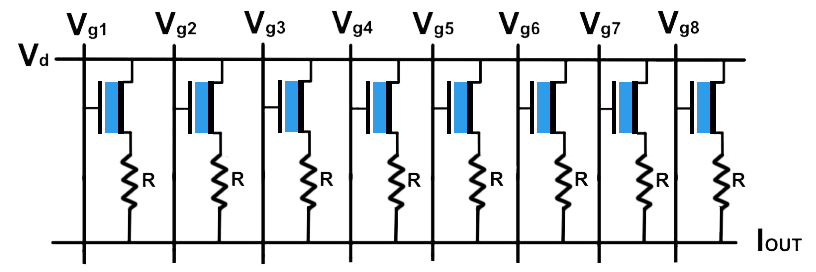}
     \vspace{-1.1em}
     \caption{The traditional 8-bit wordlength 1FeFET-1R structure\cite{Soliman_2020}.}
     \vspace{-1.1em}
 \label{fig:1FeFET1R}
 \end{figure}

To assess the significance of our proposed design, we first analyze traditional FeFET based CiM designs under different temperature conditions. As an example, we consider the basic 1FeFET-1R structure proposed in \cite{Soliman_2020}. Fig. \ref{fig:1FeFET1R} illustrates the basic 1FeFET-1R array structure. 
For our analysis, we maintain all given parameters, e.g., write voltage and latency, and adjust the read voltage to investigate both the saturation region ($V_{read}=1.3V$) and the subthreshold region ($V_{read}=0.35V$).

\begin{figure}
     \centering
     \includegraphics[width=1\linewidth]{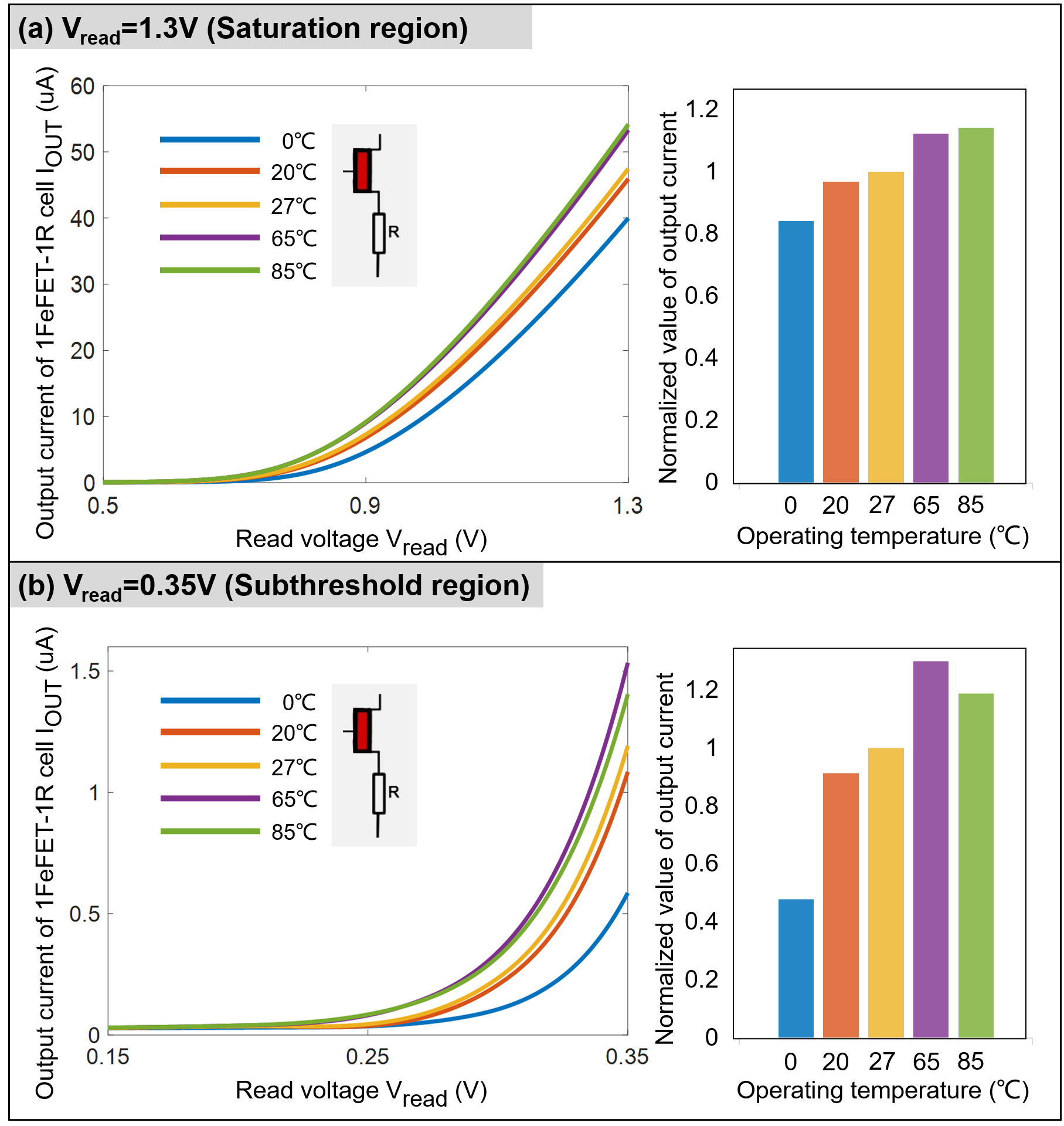}
     \vspace{-2em}
     \caption{Output current of 1FeFET-1R cell at temperatures from 0℃ to 85℃ under the same input voltages and normalized output current with reference temperature at 27℃. (a) $V_{read}=1.3V$ (saturation region, which is the operating voltage of \cite{Soliman_2020}). (b) $V_{read}=0.35V$ (subthreshold region). Current fluctuation is measured by the difference of nominal output current to 1.}
     \vspace{-1.5em}
 \label{fig:1FeFET}
 \end{figure}

\begin{figure}
     \centering
     \includegraphics[width=0.85\linewidth]{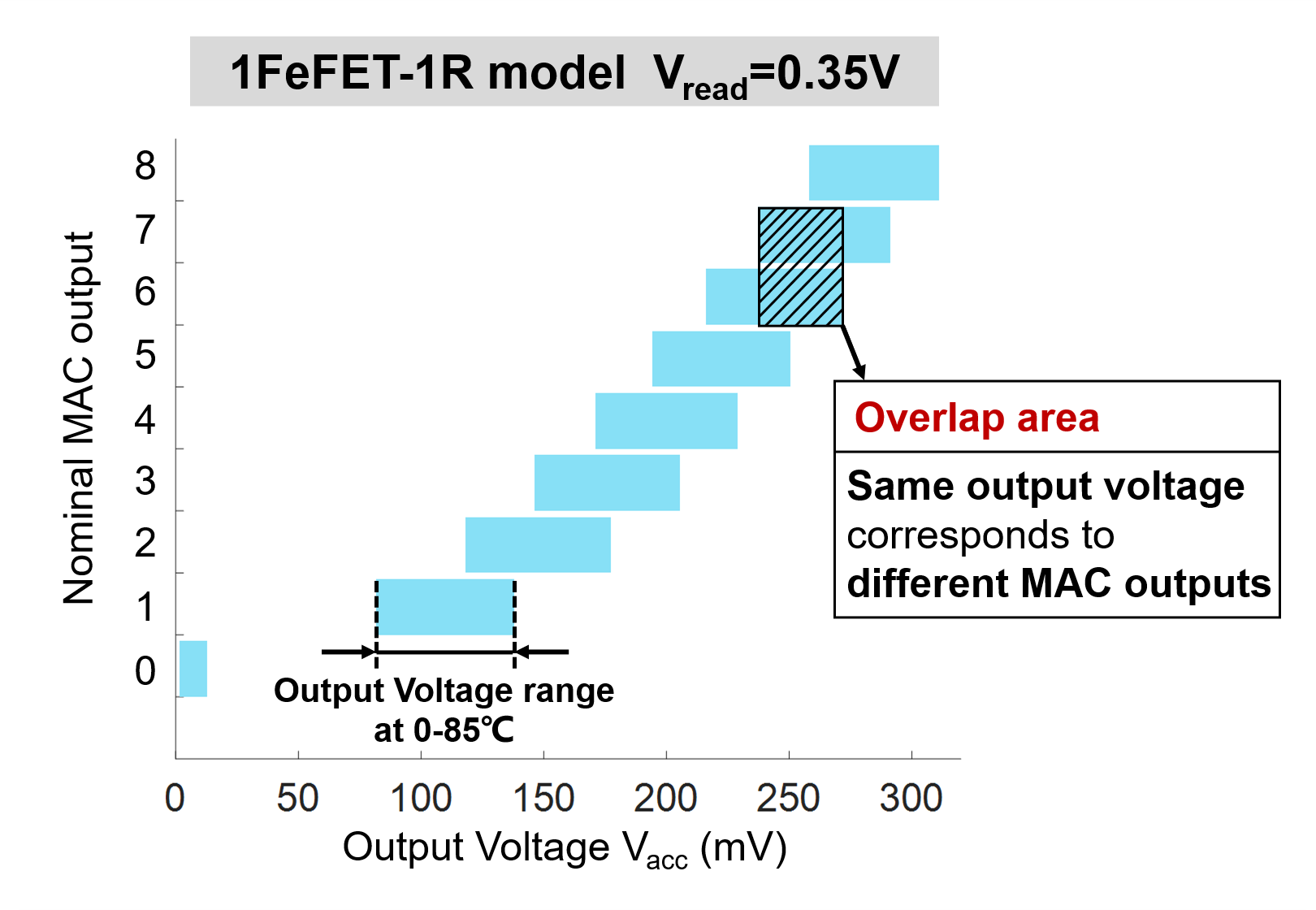}
     \vspace{-1.5em}
     \caption{Output voltage ranges at temperatures from 0℃ to 85℃ of an 1FeFET-1R CiM array with 8 cells per row. MAC outputs are from 0 to 8 and  overlap with each other, causing  computation errors.}
     \vspace{-1.2em}
     \label{fig:overlap}
 \end{figure}

First we analyze the output current of a single 1FeFET-1R cell. Fig. \ref{fig:1FeFET}(a) and (b) illustrate the output current and the normalized output current to show current fluctuations (over the reference current at 27℃) at different temperatures with the same input voltages upon read  when cells are  operating in the saturation region and subthreshold region, respectively. 
When temperature rises from 0℃ to 85℃, the fluctuation of the output current reaches up to $20.6\%$ (saturation region) and $52.1\%$ (subthreshold region), respectively. 
These large current fluctuations in output current pose challenges in constructing larger-scale CiM arrays, where different MAC outputs may have overlapped output currents due to temperature drift, thus  causing computation inaccuracies.
Fig. \ref{fig:overlap} 
displays the range of output results obtained from an 1FeFET-1R CiM array with 8 cells per row, with a read voltage $V_{read}$ of 0.35V, under varying temperatures. 
It can be seen that problems arise when two distinct MAC outputs overlap with each other due to the impact of temperature drift.

We extend our analysis to several other FeFET based structures, specifically those operating in the subthreshold region, such as the one proposed in \cite{Lu_2022}. Upon examining these works, we observe that FeFET devices operating in or near the subthreshold region demonstrate expected performance within specific temperature conditions. However, these devices exhibit erroneous results when temperature changes, highlighting the need for robust temperature compensation techniques.

\subsection{ Proposed 2T-1FeFET Design}

\begin{figure}
    \centering
    \includegraphics[width=0.9\linewidth]{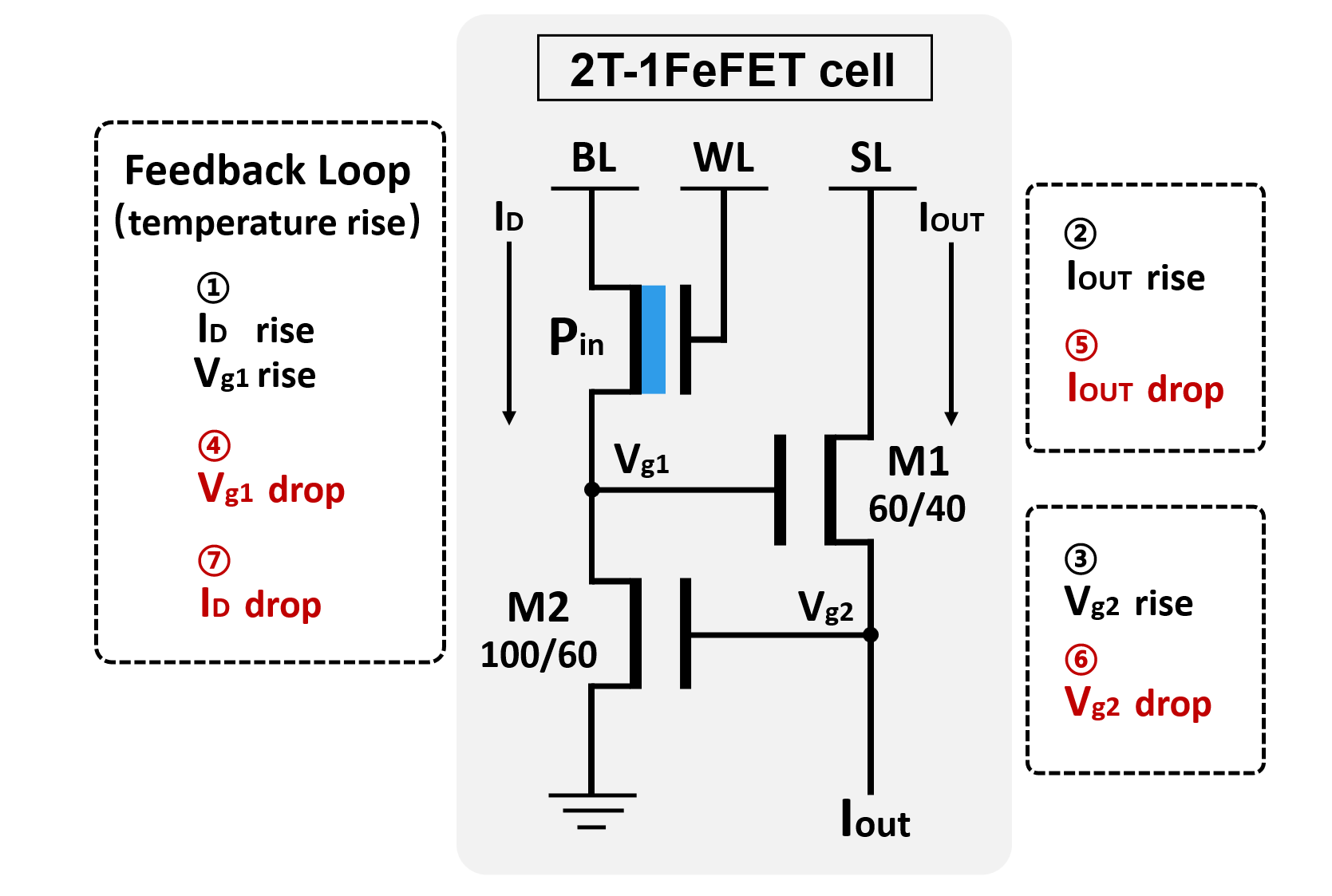}
    \vspace{-1em}
    \caption{Schematic of the proposed 2T-1FeFET cell design for CiM array, with the feedback loop for temperature drift compensation.
    }
    \vspace{-1.6em}
    \label{fig:cell}
\end{figure}

\begin{figure*}
    \centering
    \includegraphics[width=0.85\linewidth]{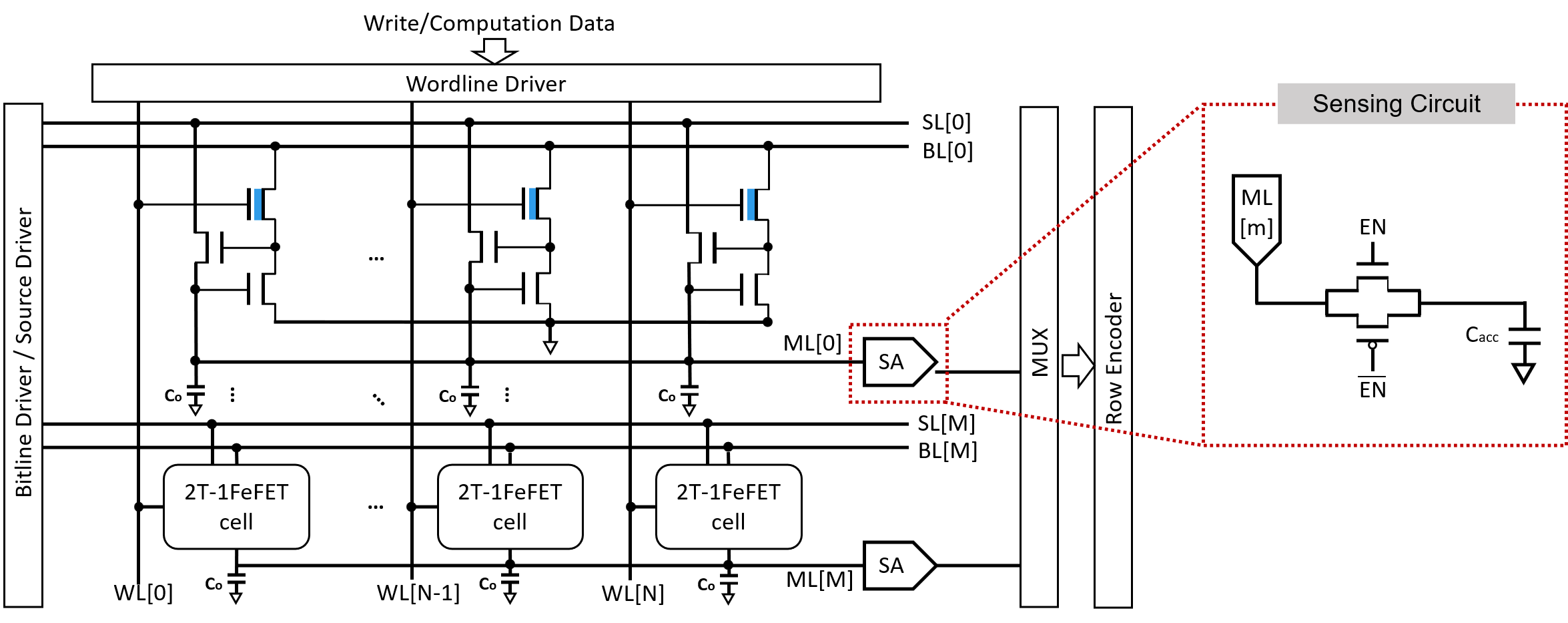}
    \vspace{-1em}
    \caption{Overall schematic of the proposed subthreshold 2T-1FeFET based CiM array and the associated sensing circuit.}
    \vspace{-1.5em}
    \label{fig:array}
\end{figure*}

To achieve temperature resilience, we 
propose a  FeFET based cell as shown in Fig. \ref{fig:cell}. In this structure, two n-MOSFETs (M1 and M2) are connected in a ring configuration.
Both M1 and M2 operate in the subthreshold region. 
The cell parameters, such as the W/L (width/length) ratio, read latencies, and write latencies, are  tuned to improve the temperature resilience of the cell.
As the temperature increases and the FeFET drain current $I_D$ rises, the gate voltage of M1 also increases due to the presence of M2. 
By keeping voltages on BL and SL lines constant, the drain current of M1 increases as a result of the increasing gate voltage. This increased drain current of M1, i.e., the output current, subsequently raises the voltage at the gate of M2. 
This leads to a decrease in $V_{gs1}$ (gate-source voltage of M1), limiting and reducing the drain current of M1, which is the output current of cell, and then leads to the voltage drop on $V_{gs2}$ (gate-source voltage of M2). Ultimately, the drain current of M2 and the FeFET is reduced. 
Similarly, the current drop of  FeFET due to decreasing temperature will also rise back per the feedback loop.
By employing this feedback mechanism, both the drain currents of M1 and M2 experience a drop/rise as temperature increases/decreases, effectively mitigating the impact of temperature drift on the cell output. 

The overall structure of the CiM array is built as shown in Fig. \ref{fig:array}. 
Each row performs one MAC operation. 
In this design, we build a  CiM array with 8 2T-1FeFET cells per row, with each cell associating with an small capacitor that converts current signals to stable voltage signals. 
The sensing circuit of array illustrated in Fig. \ref{fig:array} consists of a switch controlled by signal $EN$, and an accumulation capacitor $C_{acc}$ to obtain the output of MAC operations.

During the write operation of the 2T-1FeFET CiM array, 
we apply a -4V pulse for 200ns to  WLs  to program FeFETs to high-$V_{TH}$ state (logic ‘0’). 
On the other hand, to set the FeFETs to the low-$V_{TH}$ state (logic '1'), a +4V pulse for 115ns is applied. 
During the MAC operation, BL is set to 1.2V and SL is set to 0.2V. 
Within a row, each cell is controlled by an input WL line. When the input is '1', WL is set to 0.35V to activate  subthreshold-FeFETs  with low-$V_{TH}$ state, i.e., storing '1'.
Conversely, when the input is '0', WL disables FeFETs, conducting no currents. 
If a FeFET stores high-$V_{TH}$ state, the multiplication result of the cell remains 0 regardless of input on WL.
Each cell performs the multiplication operation based on the input voltage and the stored state, and multiplication currents are drawn from the SL lines, charging all the cell capacitance $C_{o}$s. 
After the charging, the $EN$ signal is set to a high level, and switches are opened simultaneously to charge the accumulation capacitance $C_{acc}$. The stored voltage of $C_{acc}$ can be calculated using equation (1):

\begin{equation}
V_{acc}=\frac{C_{o}}{nC_{o}+C_{acc}}\sum_{i=1}^{n}V_{Oi}   
\end{equation}

where $n$ is the number of cells connected to $C_{acc}$, and $V_{Oi}$ is the voltage level of each $C_{o}$. 
The array demonstrates resilience to temperature drift due to the temperature-resilient nature of its cells. Furthermore, the array exhibits enhanced performance as the effects of temperature drift on both the cells and the sensing circuit align in the same direction.


\section{Validation and Evaluation}

\label{sec:eval}
In this section, we first evaluate the performance of our proposed 2T-1FeFET CiM structure, including the resilience to temperature drift from 0℃ to 85℃ and process variation, and apply our design to VGG network with Cifar-10 dataset.
In the end, we compare the the proposed 2T-1FeFET CiM structure with other emerging designs.The evaluations are performed on Cadence Virtuoso Spectre simulator.

\subsection{Temperature Resilience Validation}
\begin{figure}
    \centering
    \includegraphics[width=1\linewidth]{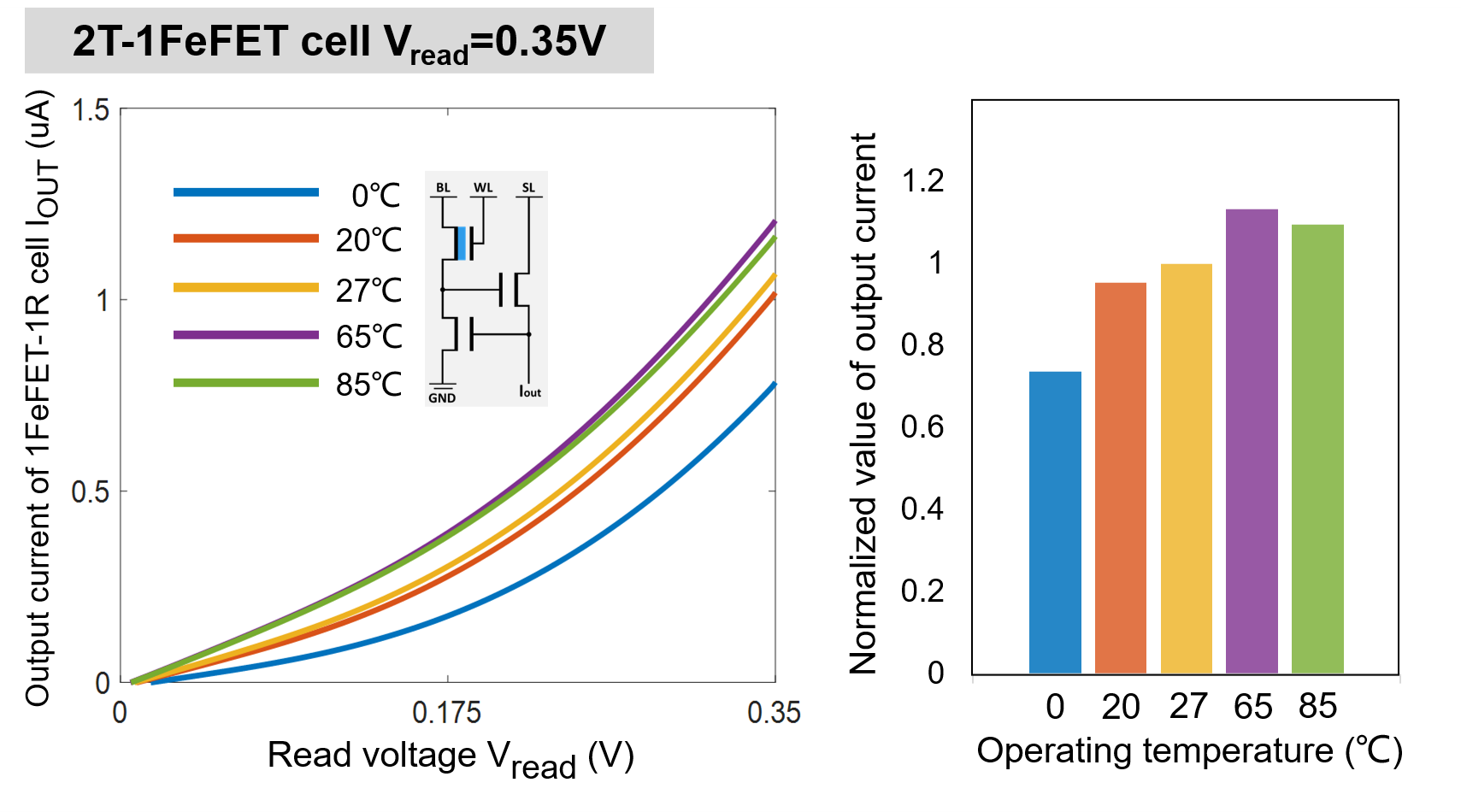}
    \vspace{-2.5em}
    \caption{Normalized output current of proposed 2T-1FeFET cell under the same input voltages with varying temperature, the reference temperature is 27℃. 
    The 2T-1FeFET cell maintains similar temperature resilience  to 1FeFET-1R structures that operate in the saturation region, and  performs even better at  temperatures above 20℃.}
    \vspace{-1em}
    \label{fig:swing}
\end{figure}

Fig. \ref{fig:swing} shows the temperature impact on our proposed cell design. The results are tested with the same input voltages and the reference temperature is 27℃ (RT). 
The largest current fluctuation over the reference temperature of 2T-1FeFET cell is $26.6\%$ at 0℃, which is close to the $20.6\%$ fluctuation of 1FeFET-1R cell that operates in the saturation region (as revealed in Sec. \ref{sec:proposedwork}), and much better than subthreshold 1FeFET-1R cell, whose current fluctuation reaches 52.1$\%$. Moreover, when temperature exceeds 20℃, our design outperforms the saturated 1FeFET-1R design, with the largest current fluctuation reduced to $12.4\%$. 

\begin{figure}
    \centering
    \vspace{0.5em}
    \includegraphics[width=1.02\linewidth]{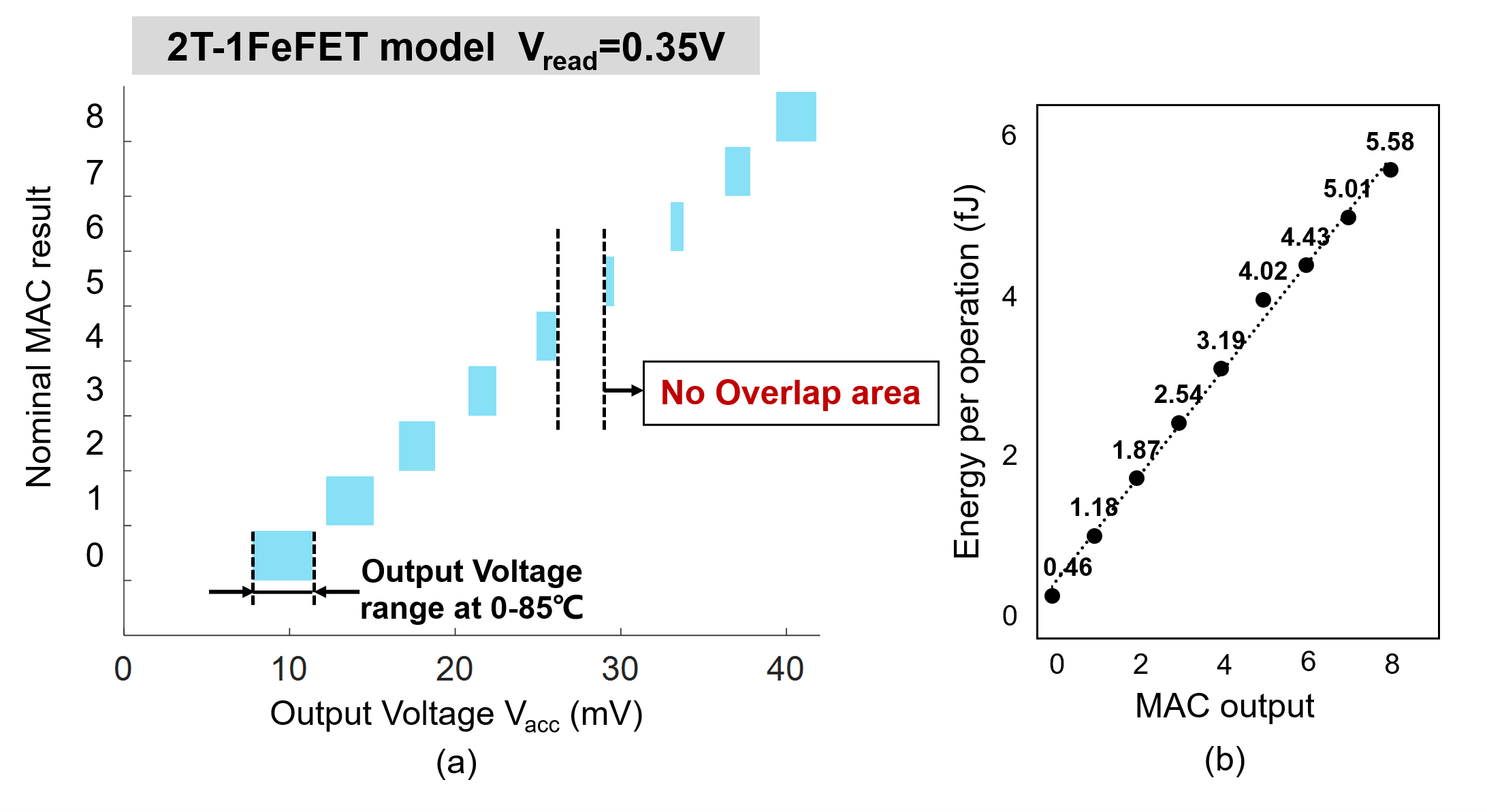}
    \vspace{-2.5em}
    \caption{
     (a) MAC output ranges of the proposed 2T-1FeFET array containing 8 cells per row with varying  temperature.  MAC outputs vary from 0 to 8. (b) Energy consumption per operation at different MAC outputs.}
    \label{fig:MAC_result}
    \vspace{-1.3em}
\end{figure}

We further  perform simulation on 2T-1FeFET CiM array with 8 cells per row to measure the different outputs of MAC operation under different temperatures. Fig. \ref{fig:MAC_result} illustrates different MAC outputs and respective energy consumption of our proposed 2T-1FeFET CiM array with 8 cells per row. 
The MAC outputs under different temperatures do not overlap, indicating stable CiM computation.
Fig. \ref{fig:MAC_result}(b) shows the energy consumption of our design corresponding to different MAC outputs.
With each MAC operation consisting of 8 multiplications and 1 accumulation, the average energy per operation is about 3.14fJ, and the energy efficiency reaches 2866 TOPS/W.

\label{sec:MAC_result}

To better evaluate our 2T-1FeFET array, we define Noise Margin Rate ($NMR$) whose numerical expression is (2):
\begin{equation}
NMR_i=\frac{LV_{i+1}-HV_i}{HV_i-LV_i}    
\end{equation}

\noindent In this expression, $HV_i$ and $LV_i$ are the highest and lowest output voltages for $MAC=i$ within 0℃ to 85℃. $NMR_i$ values the ratio of the gap distance between two different MAC outputs to the width of the $MAC=i$ output range. $NMR$ also signifies the probability of two different outcomes overlapping with each other. We further define $NMR_{min}$ expressed by equation (3):

\begin{equation}
NMR_{min}=min\{NMR_i\},i=0,1,\ldots,8
\end{equation}

\noindent to show the worst performance of the array from $MAC=0$ to $MAC=8$. A higher and positive value of $NMR_{min}$ indicates better performance of the CiM array, while a negative value suggests that the array may output overlaps and induce errors. As far as we have studied, all existing FeFET based CiM designs with 8 cells per row, whether operating in the saturation region or subthreshold region, exhibit $NMR_{min}$ values below 0  within 0℃ to 85℃.
Our array exhibits an $NMR_{min}$ value of $NMR_{min}=NMR_0=0.22$. When only considering the temperature range of 20℃ to 85℃, the NMR value increases to $NMR_{min}=NMR_7=2.3$, indicating a significant improvement in sense margin. 
It is evident that our design has better optimization for the range of 20℃ to 85℃ than the range of 0℃ to 20℃. Despite this, our design outperforms other related designs upon temperature drift, maintaining superior overall performance.



\begin{figure}
    \centering
    \includegraphics[width=0.8\linewidth]{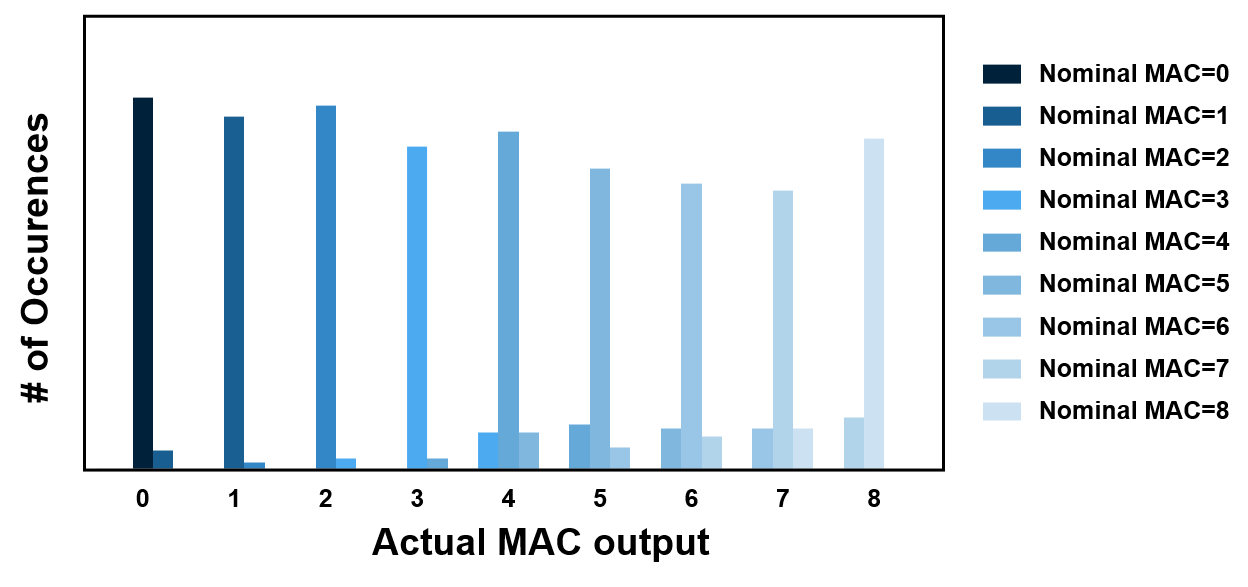}
    \vspace{-1em}
    \caption{
    The impact of process variation with $\sigma_{V_T}=54mV$ at 27℃ on CiM output via 100 Monte Carlo simulations.}
    \label{fig:error}
\end{figure}

To further assess the impact of process variations on FeFET based CiM design,
we run 100  Monte Carlo simulations of the  2T-1FeFET CiM array with an experimental FeFET Gaussian variability of $\sigma_{V_T}=54mV$.
The simulation results, shown in a histogram (Fig. \ref{fig:error}), reveal that the highest error caused by process variation is approximately $25\%$, which is not significantly higher than other emerging CiM designs, such as the 6T SRAM CiM \cite{Ali_2020} that has a maximum error of $50\%$ caused by process variation. 
Our proposed CiM array exhibits an error below $10\%$ when reduced to 4 cells per row, which is comparable to the 1FeFET-1R design \cite{Soliman_2020}. 


\label{sec:error}

\begin{table}[t!]
 \vspace{-3.2mm}
\centering
\caption{The structure of VGG Executed on Cifar-10 Dataset}
 \vspace{-3mm}
\label{tab:cifar10}
\resizebox{1\columnwidth}{!}{
\begin{tabular}{|c|c|c|c|}
\hline
Layer& Input Map & Output Map & Non Linearity \\ \hline
64 3$\times$ 3 Conv1& 32$\times$32$\times$3   & 32$\times$32$\times$64 & ReLU,dropout(0.3) \\ 
\hline
64 3$\times$ 3 Conv2& 32$\times$32$\times$64   & 32$\times$32$\times$64 & ReLU \\ 
\hline
$[2,2]$ MaxPool1 & 32$\times$32$\times$64   & 16$\times$16$\times$64& $-$  \\
\hline
128 3$\times$ 3 Conv3& 16$\times$16$\times$64   & 16$\times$16$\times$128 & ReLU,dropout(0.4) \\ 
\hline
128 3$\times$ 3 Conv4& 16$\times$16$\times$128   & 16$\times$16$\times$128 & ReLU \\ 
\hline
$[2,2]$ MaxPool2 & 16$\times$16$\times$128 & 8$\times$8$\times$128& $-$  \\ 
\hline
256 3$\times$ 3 Conv5& 8$\times$8$\times$128   & 8$\times$8$\times$256 & ReLU,dropout(0.4) \\ 
\hline
256 3$\times$ 3 Conv6& 8$\times$8$\times$256   & 8$\times$8$\times$256 & ReLU,dropout(0.4) \\ 
\hline
256 3$\times$ 3 Conv7& 8$\times$8$\times$256   & 8$\times$8$\times$256 & ReLU \\ 
\hline
$[2,2]$ MaxPool3 & 8$\times$8$\times$256 & 4$\times$4$\times$256 & $-$  \\ 
\hline
4096$\times$4096 FC1 & 1$\times$1$\times$4096 & 1$\times$1$\times$4096 & ReLU,dropout(0.5)  \\ 
\hline
4096$\times$4096 FC2 & 1$\times$1$\times$4096 & 1$\times$1$\times$4096 & ReLU,dropout(0.5)  \\ 
\hline
4096$\times$10 FC3 & 1$\times$1$\times$4096 & 1$\times$1$\times$10 & $-$ \\ 
\hline

\end{tabular}
}
 \vspace{-2em}
\end{table}

\begin{table*}
\fontsize{8}{10}
 \vspace{-5mm}
\centering
\caption{Performance Summary}
 \vspace{-3mm}
\label{tab:result}
\renewcommand{\arraystretch}{1.4}
\begin{tabular}{c|c|c|c|c|c|c|c|c}
\toprule
\textbf{\shortstack{Related\\ Work}}& \textbf{Device}&\textbf{Process}& \textbf{\shortstack{Cell \\Structure}} & \textbf{Dataset}&\textbf{\shortstack{Network \\Architecture}}& \textbf{Accuracy}& \textbf{Energy} &\textbf{\shortstack{Energy \\Efficiency}}  \\ 
\toprule
\cite{Ali_2020}& CMOS& 65nm &6T SRAM&\shortstack{Cifar-10\\MNIST} & \shortstack{VGG\\LeNet-5} &\shortstack{88.83$\%$\\99.05$\%$}& \shortstack{NA\\158.203nJ (/inference)} & NA  \\ 
\hline
\cite{Yin_2020}& CMOS&65nm& 12T SRAM& Cifar-10&BNN& 85.7$\%$& 2.48-7.19fJ (/operation) & 403 TOPS/W\\ 
\hline
\cite{Soliman_2020}& FeFET&28nm & 1FeFET-1R&/&/&/& NA &13714 TOPS/W \\ 
\hline
\cite{Sk_2023}& FeFET&28nm & 1FeFET-1T&MNIST&MLP&97.6$\%$& 17.6uJ (/inference) &NA \\ 
\hline
\cite{Yu_2021}& ReRAM &22nm & 1T-1R&Cifar-10&VGG&91.72$\%$& $\approx$5.5uJ (/inference) &26.66 TOPS/W \\ 
\hline
\cite{Jacob_2023}& MTJ&28nm & 1T-1MTJ&/&/&/& 1.4pJ (/operation) &32 TOPS/W \\ 
\hline
This Work & FeFET & 14nm&2T-1FeFET&Cifar-10& VGG & 89.45$\%$&\shortstack{\\85.08nJ (/inference)\\3.14fJ (/operation) }&2866 TOPS/W \\ 
\toprule

\end{tabular}
 \vspace{-2.1em}
\end{table*}

\subsection{Performance Evaluation}


In this section, we  evaluate the performance of the proposed design in the context of convolutional neural network (CNN) layers. 
Here we use the same CNNs and datasets as in \cite{Ali_2020}, i.e., VGG and Cifar-10 dataset. 
The executed VGG network is summarized in Table \ref{tab:cifar10}, and the performance of the 2T-1FeFET cell applied in the VGG layers is illustrated in Table \ref{tab:result}.
Monte Carlo simulations of executing VGG on Cifar-10 using the proposed hardware suggest an average classification accuracy of $89.45\%$. 

we  compare our subthreshold-FeFET based CiM design with other existing CiM designs, which are SRAM CiM designs \cite{Ali_2020,Yin_2020}, ReRAM CiM designs \cite{Yu_2021} and MTJ CiM designs \cite{Jacob_2023}. 
As shown in Table \ref{tab:result}, we list several performance metrics of our design and other types of existing CiM designs. 
The 2T-1FeFET design demonstrates significantly lower power consumption compared to other energy-efficient designs, thanks to its high $I_{ON}$/$I_{OFF}$ ratio, and subthreshold computing mode. 
Other designs such as ReRAM and MTJ consume 64.6$\times$ and 445.9$\times$ more operation energy than 2T-1FeFET CiM array. This highlights the advantage on power consumption of subthreshold-FeFET based CiM designs over other designs.
The proposed design has a latency of 6.9 ns for each MAC operation. While it may not be as fast as some other devices like the 1FeFET-1R, it remains competitively efficient. 
The lower operating voltage and accumulative capacitors contribute to the slightly higher latency.

From all these comparisons, we conclude that our 2T-1FeFET design has significant advantages on power consumption while being resilient to temperature variations, and the accuracy of our design applied in VGG network is relatively high. 
Notably, We are  the first to realize the temperature-resilient subthreshold-FeFET based CiM array with 8 cells per row within 0℃ to 85℃ range, especially at temperature above 20℃, whose property other existing designs do not possess.


\section{Conclusion}
\label{sec:conclusion}
In this paper, to build ultra-low power CiM design for practical edge scenarios,
we investigate the computation failure caused by thermal variations of FeFET based CiM structures, and  propose a novel subthreshold 2T-1FeFET cell design and an ultra-low power CiM array with 8 cells per row that are immune to output overlap induced operation failure at temperatures ranging from 0℃ to 85℃. The developed FeFET based CiM design has a remarkable reduction on power consumption compared to other emerging designs, and a remarkable energy efficiency of averaging 2866 TOPS/W for the CiM array with 8 cells per row. 
Evaluation also shows $89.45\%$ accuracy of the proposed design on the VGG neural network architectures running the CIFAR-10 dataset. Overall, our proposed CiM structure offers a promising solution to mitigate the impact of temperature variations, reduce power consumption, and deliver reliable MAC operations for neural networks deployed in edges.


\section*{Acknowledgements}
This work was supported in part by  National Key R\&D Program of China (2020YFB1313501), Zhejiang Provincial Natural Science Foundation (LD21F040003, LQ21F040006), NSFC (62104213, 92164203), National Key Research and Development Program of China (2022YFB4400300).

\scriptsize{
    \bibliographystyle{ieeetr}
    \bibliography{bib}
}
\end{document}